\documentclass[aps,showpacs,preprintnumbers]{revtex4}

\usepackage{graphicx}% Include figure files
\usepackage{dcolumn}% Align table columns on decimal point
\usepackage{bm}% bold math
\usepackage{epsfig}

\begin{document}

\title{The properties of kaonic nuclei in relativistic mean-field theory}
\author{
 X.\ H.\ Zhong,$^{1,2}$\footnote{E-mail: zhongxianhui@mail.nankai.edu.cn}
 G.\ X.\ Peng,$^{2}$\footnote{E-mail: gxpeng@ihep.ac.cn }
 L.\ Li,$^1$ and
 P.\ Z.\ Ning,$^{1,2}$\footnote {E-mail: ningpz@nankai.edu.cn}
       }
\affiliation{
 $^1$Department of Physics, Nankai University, Tianjin 300071, China\\
 $^2$Institute of High Energy Physics,
       Chinese Academy of Sciences, Beijing 100039, China
            }
%\date{\today}

\begin{abstract}
The static properties of some possible light and moderate kaonic
nuclei, from C to Ti, are studied in the relativistic mean-field
theory. The 1s and 1p state binding energies of $K^-$ are in the
range of $73\sim 96$ MeV and $22\sim 63$ MeV, respectively. The
binding energies of 1p states increase monotonically with the
nucleon number A. The upper limit of the widths are about $42\pm 14$
MeV for the 1s states, and about $71\pm 10$ MeV for the 1p states.
The lower limit of the widths are about $12\pm 4$ MeV for the 1s
states, and $21\pm 3$ MeV for the 1p states. If $V_{0}\leq 30$ MeV,
the discrete $K^-$ bound states should be identified in experiment.
The shrinkage effect is found in the possible kaonic nuclei. The
interior nuclear density increases obviously, the densest center
density is about $2.1\rho_{0}$.
\end{abstract}

%\keywords{kaonic nuclei, relativistic mean field, shrinkage effect}

\pacs{21.10.Dr, 21.30.Fe, 21.90.+f}

\maketitle

\section{Introduction}

The $K$-nucleon ($KN$) interactions in nuclear matter have been an
interesting topic in nuclear physics, since Kaplan firstly discussed
the question of kaon condensation in dense nuclear matter about
twenty years ago \cite{Kaplan}. With several typical methods, such
as chiral perturbation theory (ChPT) \cite{c5,brown,c4}, RMF model
\cite{c1,c4,c5,c3}, chiral coupled channel model \cite{Weis,d4},
chiral unitary model \cite{oset,Hirenzaki} or other chiral model
\cite{a2,a3}, and the phenomenological model by fitting the $K^{-}$
atomic data using a density-dependent optical potential (DD model)
\cite{Frid,Frid1999}, a strong attractive $K^{-}$-nucleus potential
at threshold is predicted. The $K^{-}$-nucleus potential strongly
depends on the model used. The DD model gives the deepest inner
$K^{-}$-nucleus potential which is in the range of $150\sim200$ MeV
\cite{Frid,Frid1999}. The chiral coupled channel model predicts the
$K^{-}$-nucleus potential in the range of $85\sim 120$ MeV
\cite{Weis,c3,c4}, close to that by RMF model (The $\sigma$-$K$
coupling constant determined by the $KN$ scatering length). The
chiral models give shallower $K^{-}$-nucleus potential $50\sim 70$
MeV \cite{c4,oset,Hirenzaki,a2,a3}.

Based on the strong attractive $K^{-}$-nucleus potential, there
maybe exist ``deeply bound kaonic atoms'' or kaonic nuclei. The
existence of the ``deeply bound kaonic atoms'', or, kaonic nuclei
was already well-known in later 1990s \cite{deepbound, a4}. It was
suggested that the kaonic nuclei could be produced by the ($K^-$,
$p$) and ($K^-$, $n$) reactions \cite{a4}. Since then, the issue on
the $K^-$-nucleus deeply bound states (kaonic nuclei) has been a hot
topic.

Theoretically, there have been many works on kaonic nuclei. The
binding energies, the widths, the production methods, the nuclear
densities around $K^-$-meson, and other aspects were discussed with
different models
\cite{a2,a6,a3,a8,a9,a10,a11,d,Galaa,Agnello,Dote,Yamagata,OsetToki}.
Experimentally, some evidences for the candidate states in the
($K_{\mathrm{stop}},\ n)$ and ($K_{\mathrm{stop}},\ p)$ reactions on
$^4$He \cite{b1,b2}, and in the ($K^-$, $n$) in-flight reaction on
$^{16}$O \cite{b3} are reported. Recently, by using the FINUDA
spectrometer installed at the e$^+$e$^-$ collider DA$\Phi$NE, the
FINUDA collaboration succeeded in detecting a $K^-$ bound state
$K^-pp$ through its two-body decay into a $\Lambda$\ hyperon and a
proton \cite{b4}. However, just some months ago, Magas \emph{et al.}
reanalyzed the FINUDA data and claimed that the experimental
spectrum can be naturally explained in their Monte Carlo simulation
of the $K^-$ absorption events in nuclei, without the need of exotic
mechanisms like the formation of a $K^-pp$ bound state \cite{b5}.
Thus, further experimental and theoretical works are needed on this
issue.

In the present work, we will study the static properties of the
possible kaonic nuclei in the framework of relativistic mean-field
theory (RMF). The pioneer works, describing the in-medium $K^-N $
interaction with RMF method, can be found in  \cite{c1,c2,c3}. Since
then, the RMF model had been widely used to describe the in-medium
properties of (anti)kaon mesons \cite{c4,c5}, the $K^-$-nucleus
interaction
 \cite{Frid1999}, the $K^-$-nucleus elastic scattering \cite{c7} and
the kaonic nuclei \cite{d,Galaa}. In this work, the static
properties of kaonic nuclei are obtained by solving,
self-consistently, the equations of motion for nucleons and mesons,
which are derived from the Lagrangian for nucleons and (anti)kaon
mesons. To describe the widths of the $K^-$ in nuclei, the equation
of motion for $K^-$ are modified by introducing an imaginary part of
the self-energy phenomenologically. In the calculations, we will
focus on the energy spectra, the nuclear density distribution, the
r.m.s.\ radii of proton, neutron and charge distribution for the
possible kaonic nuclei.

The paper is organized as follows. In the subsequent section, the
Lagrangian density is given, the equations of motion for nucleons
and the meson fields $\sigma$, $\omega$, $\rho$, and Kaons are
deduced, the imaginary part of the self-energies are introduced and
the binding energies of $K^-$ are defined. We then present our
results and discussions of the obtained properties of kaonic nuclei
in Sec. III. Finally a summary is given in Sec. IV.

\section{Formulas}

In relativistic mean field theory, the standard Lagrangian density
for an ordinary nucleus can be written as \cite{d1,d2}
\begin{eqnarray}
{\mathcal{L}}_{0}
={\mathcal{L}}_{\mathrm{Dirac}}
 +{\mathcal{L}}_{\sigma}
 +{\mathcal{L}}_{\omega}
 +{\mathcal{L}}_{\rho}
 +{\mathcal{L}}_{A},
\end{eqnarray}
where
\begin{eqnarray}
{\mathcal{L}}_{\mathrm{Dirac}}
&=&
 \bar{\Psi}_N(i\gamma^{\mu}\partial_{\mu}-M_{N})\Psi_N, \\
{\mathcal{L}}_{\sigma}
&=&
 \frac{1}{2}\partial_{\mu}\sigma\partial^{\mu}\sigma
 -\frac{1}{2}m_{\sigma}^2\sigma^2
 -g_{\sigma N }\bar{\Psi}_N\sigma\Psi_N
\nonumber\\
& &
 -\frac{1}{3}g_{2}\sigma^3
 -\frac{1}{4}g_{3}\sigma^4,\\
{\mathcal{L}}_{\omega}
&=&
 -\frac{1}{4}F_{\mu\nu}F^{\mu\nu}
 +\frac{1}{2}m_{\omega}^2\omega_{\mu}\omega^{\mu}
\nonumber\\
& &
 -g_{\omega N}\bar{\Psi}_N\gamma^{\mu}\Psi_N\omega_{\mu}, \\
{\mathcal{L}}_{\rho}
&=&
 -\frac{1}{4}\vec{G}_{\mu\nu}\vec{G}^{\mu\nu}
 +\frac{1}{2}m_{\rho}^2
\vec{\rho}_{\mu}\cdot\vec{\rho}^{\mu}\nonumber\\
& & -g_{\rho N}\bar{\Psi}_{N}\vec{\rho}^{\mu}\cdot \vec{I}\Psi_{N},\\
{\mathcal{L}}_{A}
&=&
  -\frac{1}{4}H_{\mu\nu}H^{\mu\nu}
  -e\bar{\Psi}_N\gamma_{\mu}I_{c}A^{\mu}\Psi_N,
\end{eqnarray}
with
\begin{eqnarray}
F_{\mu\nu}
&=&
 \partial_{\nu}\omega_{\mu}-\partial_{\mu}\omega_{\nu},\\
\vec{G}_{\mu\nu} &=&
 \partial_{\nu}\vec{\rho}_{\mu}-\partial_{\mu}\vec{\rho}_{\nu}, \\
H_{\mu\nu}
&=& \partial_{\nu}A_{\mu}-\partial_{\mu}A_{\nu},
\end{eqnarray}
where the meson fields are denoted by $\sigma$, $\omega_{\mu}$,
and $\vec{\rho}_{\mu}$, and their masses by $m_{\sigma}$,
$m_{\omega}$, $m_{\rho}$, respectively.  $\Psi_N$ is the nucleon
field with corresponding mass $M_N$. $A_{\mu}$ is the
electromagnetic field. $g_{\sigma N} $, $g_{\omega N} $, and
$g_{\rho N} $ are, respectively, the $\sigma$-$N$, $\omega$-$N$,
and $\rho$-$N$ coupling constants. $I_{c}=(1+\tau_{3})/2$ is the
Coulomb interaction operator with $\tau_3$ being the third
component of the isospin Pauli matrices for nucleons. $I$ is the
nucleon isospin operator. In this paper, we adopt the NL-SH
parameter set \cite{d3}, which describes the properties of finite
nuclei reasonably. The masses and coupling constants are listed in
Tab. \ref{paramtab}.

For a $K^-$-nucleus system, another Lagrangian density
${\mathcal{L}}_{\mathrm{K}}$ describing the (anti)kaon interaction
with nucleons should be added to ${\mathcal{L}}_0$. Kaons are
incorporated into the RMF model by using the $KN$ interactions
motivated by the one-meson-exchange models \cite{d0}. In the
meson-exchange picture, the scalar and vector interactions between
kaons and nucleons are mediated by the exchange of $\sigma$ and
$\omega$ mesons, respectively.  The coupling of the (anti)kaon to
the isovector $\rho$ meson is here excluded due to $N= Z$ nuclear
cores. Thus, the simplest kaon-meson interaction Lagrangian density
${\mathcal{L}}_{\mathrm{K}}$ is written as \cite{c3}
\begin{eqnarray}
{\mathcal{L}}_{\mathrm{K}}
&=&
 \partial_{\mu}\bar{K}\partial^{\mu}K
 -m_K^2\bar{K}K-g_{\sigma K}m_{K}\bar{K}K\sigma
\nonumber\\
& &
 -ig_{\omega K}(\bar{K}\partial_{\mu}K-K\partial_{\mu}\bar{K})\omega^{\mu}
\nonumber\\
& &
 +(g_{\omega K}\omega^{\mu})^2\bar{K}K,
\end{eqnarray}
where $g_{\sigma\mathrm{K}}$ and $g_{\omega\mathrm{K}}$ are the
$\sigma$-$K$\ and $\omega$-$K$ coupling constants.
$g_{\omega\mathrm{K}}$ is chosen from the SU(3) relation assuming
ideal mixing, i.e., $2g_{\omega\mathrm{K}}=2g_{\pi\pi\rho}=6.04$.
$g_{\sigma\mathrm{K}}$ can be obtained from several methods, e.g.
the Bonn model \cite{d0}, by reproducing the strongly attractive
potential seen in kaonic atoms \cite{Frid1999} or by fitting the
$KN$ scattering lengths in experiments \cite{c3,c4}. In this paper,
we take the modest $K$-$\sigma$ coupling, $g_{\sigma K}  =2.088
(\approx g_{\sigma N}/5)$, by fitting the experimental $KN$
scattering length \cite{c4}. The advantages of this method were
pointed out in \cite{c4}. With these determined coupling constants,
the antikaon optical potential at normal nuclear density is about
$U_{\mathrm{K}^-}=-(85\sim 100)$ MeV \cite{c3}, which is compatible
with several groups' predictions. For example, Akaishi \emph{et al.}
gave $-119$ MeV for a $K^-$ in nuclear matter at the normal density
\cite{a9}, and Weise's gave $U_{\mathrm{K}^-}=-(120\sim 130)$ MeV by
using chiral $KN$ interaction \cite{d4}. It is interesting that the
recent experiment predicts the in-medium $K^-N$ potential is on the
order of $-80$ MeV at normal nuclear density \cite{Sch2006}, which
is also agreement with the $K^-N$ potential adopted in present
paper.

\begin{table}[ht]
\begin{center}
\caption{Parameters used in the present calculations.}
\label{paramtab}
\begin{tabular}{|cccc|cccc|}\hline\hline
\multicolumn{4}{|c|}{masses (MeV)}
     & \multicolumn{4}{|c|}{couplings}  \\ \cline{2-3}\cline{6-7}
$M_N$ & $m_{\sigma}$ & $m_{\omega}$ & $m_{\rho}$
                 & $g_{\sigma N} $ & $g_{\omega N} $
                 & $g_{\rho N} $   & $g_3$ \\ \hline
939.0 & 526.059 & 783.0 & 763.0 & 10.444 & 12.945 & 8.766 & -15.8337 \\
\multicolumn{4}{|c|}{\mbox{}}
    & \multicolumn{4}{|c|}{$g_2=-6.9099$ fm$^{-1}$}  \\ \hline\hline
\end{tabular}
\end{center}
\end{table}

In the mean field approximation, the meson-fields $\sigma$,
$\omega_{\mu}$, and $\rho_{\mu}$, and the photons $A_{\mu}$ are
replaced with their mean values, $\langle \sigma \rangle$, $\langle
\omega_{\mu}\rangle$, $\langle\rho_{\mu}\rangle$ and $\langle
A_{\mu}\rangle$, respectively. For a spherical nucleus, only the
mean values of the time components $\langle \omega_{0}\rangle$,
$\langle\rho_{0}\rangle$ and $\langle A_{0}\rangle$ remain, which
are denoted by $\omega_{0}$, and $\rho_{0}$, and $A_{0}$
respectively. Then the equations of motion for nucleons, $\omega$,
$\sigma$, $\rho$, and photons are

\begin{widetext}
\begin{eqnarray}
& \left[-i\vec{\alpha}\cdot \vec{\nabla}+\beta(M_{N}+g_{\sigma
N}\sigma_{0})+g_{\omega N}\omega_{0} +g_{\rho N}\tau_3\rho_{0}+e
I_{c}A_{0}\right]\Psi_{N}=\epsilon \Psi_{N},
&\label{eossimp1}\\
&\left(-\nabla^2+m_{\sigma}^2\right)\sigma_{0}= -g_{\sigma N}
\bar{\Psi}_{N}\Psi_{N}
-g_{2}\sigma_{0}^2-g_{3}\sigma_{0}^3-g_{\sigma K}m_{K} \bar{K}K ,
&\label{eossimp2}\\
& \left(-\nabla^2+m_{\omega}^2\right)\omega_{0}=g_{\omega N}
\bar{\Psi}_{N}\gamma^{0} \Psi_{N} -2g_{\omega K} (E+g_{\omega
K}\omega_{0})\bar{K}K ,
&  \label{eossimp3}\\
& \left(-\nabla^2+m_{\rho}^2\right)\rho_{0}=g_{\rho N}
\bar{\Psi}_{N}\gamma^{0}I \Psi_{N} ,
& \label{eossimp4}\\
& -\nabla^2A_0=e \bar{\Psi}_{N}\gamma^{0}I_{c} \Psi_{N}, &
\label{eossimp5}
\end{eqnarray}
\end{widetext}
and the equation of motion for antikaon is
\begin{eqnarray}
\left[-\nabla^2+(m_{\mathrm{K}}^2-E^2)+\Pi \right] \bar{K} =0,
\label{eosfin5}
\end{eqnarray}
with the antikaon self-energy in nuclei
\begin{eqnarray}
\Pi =-2g_{\omega\mathrm{K}}E\omega_0
+g_{\sigma\mathrm{K}}m_{\mathrm{K}}\sigma_{0}
-(g_{\omega\mathrm{K}}\omega_0)^2. \label{self}
\end{eqnarray}
In the above equations, $\epsilon$ is the nucleon single-particle
energy, and $E$ is the single-particle energy for antikaon meson. We
can see that the antikaon meson does not relate the Dirac equation
for nucleons directly. However, the presence of antikaon leads to
additional source terms in the equations of motion for meson fields
of $\sigma$ and $\omega$. Thus, the antikaon meson affects the
nucleon fields by changing the strength of meson fields of $\sigma$
and $\omega$ indirectly.

So far, we ignored the antikaon meson absorption in the nucleus,
which requires a complex potential. Within the framework of RMF
model, we are not able to calculate the imaginary part of the
potential directly. In order to include the effects of the antikaon
meson absorption in the nucleus on the calculations, and make a more
realistic estimate for the calculated results, we assume a specific
form for the antikaon self-energy with an imaginary part
\begin{eqnarray}
\widetilde{\Pi} &=&\left[-2g_{\omega\mathrm{K}}\omega_0\mathrm{Re}E
+g_{\sigma\mathrm{K}}m_{\mathrm{K}}\sigma_{0}
-(g_{\omega\mathrm{K}}\omega_0)^2\right] \nonumber \\
&& +i\left[-2(\mathrm{Re}E)f V_{0} \frac{\rho}{\rho_0}\right].
\label{selfw}
\end{eqnarray}
The similar method is also used to analyze the widths of mesonic
nuclei in \cite{d,Ludh}.

In this work, the imaginary part of the potential $\mathrm{Im}U$ is
given with the simple ``$t\rho$" form, namely,
$\mathrm{Im}U=-V_{0}\rho/\rho_0$. Where $V_{0}$ is the imaginary
potential depth at normal nuclear density, which strongly depends on
the model adopted.  The largest value $V_{0} \sim 50$ MeV is given
by fitting the experimental data of the kaonic
atoms\cite{Frid,Frid1999}, and the chiral model (only the antikaons
are dressed self-consistently) \cite{01im}. While, by fitting the
experimental data of the antikaon-nucleus scattering, the imaginary
potential depth become much shallower: $V_{0}\sim 35$ MeV \cite{c7}.
Another much shallower imaginary potential depth $20\sim 25$ MeV is
predicted by Ramos \emph{et al.} with the meson-exchange model. The
shallowest imaginary potential depth, $\sim 15$ MeV, is given by the
chiral model which incorporates the dressing of the pion and
antikaons \cite{01im}. Thus, in the present work, we set the
imaginary potential depth $V_{0}$ in the range of $15\sim 50$ MeV.

On the other hand, the phase space available for the decay
products should be reduced for deeply bound states, which will
decrease the imaginary potentials (widths). Thus, a suppression
factor, $f$, multiplying $\mathrm{Im}U$ were suggested to be
introduced by Mare\v{s} \emph{et al.} \cite{d,Galaa}. In our
calculations, we adopt their method as well. In this method, two
decay channels are considered. One is the mesonic decay channel,
$\bar{K} N\rightarrow \pi \Sigma,\ \ \pi \Lambda$. The
corresponding suppression factor is given by
\begin{widetext}
\begin{eqnarray}
f_1=\frac{M_{01}^3}{M_1^3}\sqrt{\frac{[M_1^2-(m_{\pi}+M_Y)^2][M_1^2-(m_{\pi}-M_Y)^2]}
{[M_{01}^2-(m_{\pi}+M_Y)^2][M_{01}^2-(m_{\pi}-M_Y)^2]}}
\Theta(M1-m_{\pi}-M_Y),
\end{eqnarray}
\end{widetext}
where $M_{01}=m_k+M_N$, $M_{1}=\mathrm{Re}E+M_N$ and
$Y=\Sigma,\ \Lambda$. The other channel is the non-mesonic decay
channel, $\bar{K} NN\rightarrow YN$, and the corresponding
suppression factor is
\begin{widetext}
\begin{eqnarray}
f_2=\frac{M_{02}^3}{M_2^3}\sqrt{\frac{[M_2^2-(M_{N}+M_Y)^2][M_2^2-(M_{N}-M_Y)^2]}
{[M_{02}^2-(M_{N}+M_Y)^2][M_{02}^2-(M_{N}-M_Y)^2]}}\Theta
(M_2-M_{N}-M_Y),
\end{eqnarray}
\end{widetext}
where $M_{02}=m_k+2M_N$, $M_{2}=\mathrm{Re}E+2M_N$. Since $\Sigma$
final states dominate both the mesonic  and non-mesonic decay
channels \cite{chann}, in the calculations, the hyperon $Y$ is set
as $Y=\Sigma$. The suppression factor $f$ can be assumed a mixture
of 80\% mesonic decay and 20\% non-mesonic decay \cite{chann}, thus
\begin{eqnarray}
f=0.8 f_1+0.2f_2.
\end{eqnarray}

In the calculations, in order to include the effects from the
``imaginary potential", we use the modified Klein-Gordon equation
\begin{eqnarray}
\left[-\nabla^2+(m_{\mathrm{K}}^2-E^2)+\widetilde{\Pi}  \right]
\bar{K} =0. \label{mdf}
\end{eqnarray}
The complex eigenenergies are
\begin{eqnarray}
E=-B^{s,p}_K+m_K-i\Gamma/2,
\end{eqnarray}
where the real part corresponds to the single-particle $K^-$ binding
energy, which is defined as
\begin{eqnarray}
B^{s,p}_{\mathrm{K}}=m_{\mathrm{K}}-\mathrm{Re}E,
\end{eqnarray}
and the imaginary part of the complex eigenenergies corresponds to
the widths
\begin{eqnarray}
\Gamma=-2\mathrm{Im} E.
\end{eqnarray}

The binding energy of $K^-$, $B_{K^{-}}$, is defined as the
difference between the total binding energy of the $K^-$-nucleus
system, $B(^{A}ZK^{-})$, and the total binding energy of the
ordinary nucleus, $B(^{A}Z)$ \cite{Galaa}:
\begin{eqnarray}
B_{\mathrm{K}}=B(^{A}ZK^{-})-B(^{A}Z).
\end{eqnarray}
In RMF, from the Lagrangian density (Eqs.(1---6)) we can deduce the
energy-momentum tensor $T^{\mu\nu}$ and then obtain the total
binding energy of the ordinary nucleus at once:
\begin{eqnarray}\label{hdd}
B(^{A}Z)&=&A\times
M_{N}+E_{c.m.}-\sum_{i=1}^{A}\epsilon_i\\\nonumber
&&-\frac{1}{2}\int d\textbf{r}\bigg\{-g_{\sigma
N}\sigma_0\bar{\Psi}_{N}\Psi_{N} -g_{2}\sigma_{0}^2 \\ \nonumber
&&-g_{3}\sigma_{0}^3-g_{\omega N}\omega_0
\Psi^{\dagger}_{N}\Psi_{N}\\ \nonumber
&&-g_{\rho N}
\bar{\Psi}_{N}\gamma^{0}I \Psi_{N}-e \bar{\Psi}_{N}\gamma^{0}I_{c}
\Psi_{N}\bigg\},
\end{eqnarray}
where $\epsilon_i$ is the nucleon single-particle energy labelled
$i$, and $E_{c.m.}=3/4 \cdot 41A^{1/3}$ is the center-of-mass
energy. Using the same method, according to the Eqs.(1---6) and
(10) we obtain the total binding energy of the $K^-$-nucleus
system:
\begin{eqnarray}\label{hkdd}
B(^{A}ZK^{-})&=&A\times
M_{N}+E_{c.m.}+B^{s,p}_{\mathrm{K}}\\
\nonumber &&-\sum_{i=1}^{A}\epsilon_i-\frac{1}{2}\int
d\textbf{r}\bigg\{-g_{\sigma N}\sigma_0\bar{\Psi}_{N}\Psi_{N}
\\
\nonumber &&-g_{2}\sigma_{0}^2-g_{3}\sigma_{0}^3 -g_{\omega N}\omega_0\Psi^{\dagger}_{N}\Psi_{N}\\
\nonumber &&
-g_{\rho N} \bar{\Psi}_{N}\gamma^{0}I \Psi_{N}-e \bar{\Psi}_{N}\gamma^{0}I_{c} \Psi_{N} \\
\nonumber &&+2g_{\omega K}\omega_0(3\mathrm{Re} E+g_{\omega
K}\omega_0)\bar{K}K \\
\nonumber &&-g_{\sigma K}m_{K}\sigma_{0}\bar{K}K \bigg\}.
\end{eqnarray}
From the above, we can see that for the appearance of the antikaon
there are additional terms, $-g_{\sigma K}m_{K}\sigma_{0}\bar{K}K$
and $2g_{\omega K}(3\mathrm{Re} E+g_{\omega K}\omega_0)\omega_0$, in
the integral of Eq.(\ref{hkdd}) compared with those of
Eq.(\ref{hdd}).

Solving the equations (\ref{eossimp1})--- (\ref{eossimp5}) and
Eq.(\ref{mdf}) self-consistently, we can obtain the properties of
$K^{-}$-nuclei. One point must be noted that the antikaon energy $E$
in Eq.(\ref{eossimp3}) should be replaced with its real part,
$\mathrm{Re} E$, in the calculations.

\section{Results and discussions}

The main purpose of the present calculations is to obtain the
properties of $K^-$ nuclei in RMF, such as the binding energies, the
nuclear density distributions and the r.m.s.\ radii of proton,
neutron and charge distributions. We carry out calculations for the
$N=Z$ even-even nuclear cores $^{12}$C$K^-$, $^{16}$O$K^-$,
$^{20}$Ne$K^-$, $^{24}$Mg$K^-$, $^{28}$Si$K^-$, $^{32}$S$K^-$,
$^{36}$Ar$K^-$, $^{40}$Ca$K^-$ and $^{44}$Ti$K^-$, respectively. For
the imaginary potential depths at normal nuclear density are not
well determined, in the calculations, we choose three values,
$V_{0}=15$, $30$ and $50$ MeV, respectively. The results are shown
in Tab.\ \ref{Calll}, \ref{Cal} and Figs.\ \ref{Dens}.

\begin{widetext}
\begin{center}
\begin{table}[ht]
 \caption{The single-particle $K^-$ binding energies, $B^{s,p}_{K^-}=m_K-ReE$; the binding energy of $K^-$,
 $B_{K^{-}}$; the total binding energy of the $K^-$-nucleus system,
 $B(^{A}ZK^{-})$; the total binding energy of the ordinary nucleus, $B(^{A}Z)$
 and the widths, $\Gamma$, (all in MeV), in various nuclei, where the
 complex eigenenergies are, $E=-B^{s,p}_K+m_K-i\Gamma/2$.}\label{Calll}
\begin{tabular}{ccccccc|cccc|cccccc}  \hline \hline
 &&&\multicolumn{3}{c}{\underline{$V_{0}=15$ (MeV)}}& &\multicolumn{3}{c}{\underline{$V_{0}=30$ (MeV)}}& &\multicolumn{3}{c}{\underline{$V_{0}=50$ (MeV)}}
 \\ %\cline{3-6}
&&$B(^AZ)$& $B(^AZK^-)$ & $B_K$\ \ \ &$B^{s,p}_K$& $\Gamma$ &
$B(^{A}ZK^-)$ & $B_K$\ \ & $B^{s,p}_K$ &$\Gamma$ &$B(^{A}ZK^-)$ &
$B_K$ \ \ & $B^{s,p}_K$ & $\Gamma$
\\ \hline \hline
 \\
$^{12}$C$K^-$    &$1s$& 89.6   &177.6 &\textbf{88.0} \ \ &102.3\ \ &11.2   &177.0&\textbf{87.4}  \ \ &100.4\ \  &19.4   &175.6&\textbf{86.0}    \ \  &97.6 \ \  &44.0&\\
                 &$1p$&        &115.2 &\textbf{25.6} \ \ &35.1 \ \ &23.5   &114.5&\textbf{24.8}  \ \ &33.2 \ \  &47.0    &111.2&\textbf{21.6}   \ \  &28.8 \ \  &80.9&\\
\\
$^{16}$O$K^{-}$  &$1s$& 128.5  &203.8 &\textbf{75.3} \ \ &88.3 \ \ &16.2   &203.4&\textbf{74.8}  \ \ &86.7 \ \   &32.8    &202.4 & \textbf{73.9}\ \  &84.4 \ \  &56.4\\
                 &$1p$&        &158.7 &\textbf{30.2} \ \ &41.1 \ \ &21.1   &157.7&\textbf{29.2}  \ \ &38.5 \ \   &43.0    &156.9 & \textbf{28.4}\ \  &36.6 \ \  &72.8\\
\\
$^{20}$Ne$K^{-}$ &$1s$& 142.6  &221.0 &\textbf{78.4} \ \ &90.4 \ \ &14.8    &220.8& \textbf{78.2}\ \ &89.4 \ \    &30.0    &220.0& \textbf{77.4}\ \  &87.7 \ \  &51.4\\
                 &$1p$&        &180.2 &\textbf{37.7} \ \ &48.9 \ \ &21.5    &179.9& \textbf{37.3}\ \ &47.5 \ \    &42.8    &178.5& \textbf{35.9}\ \  &45.3 \ \  &71.5\\
\\
$^{24}$Mg$K^{-}$ &$1s$&182.8   &267.5 &\textbf{84.7} \ \ &95.3 \ \ &12.2    &267.4& \textbf{84.6}\ \ &94.7 \ \   &25.2    &266.8& \textbf{84.0} \ \  &93.3 \ \  &43.6\\
                 &$1p$&        &231.0 &\textbf{48.2} \ \ &59.4 \ \ &22.0    &230.7& \textbf{47.9}\ \ &58.5 \ \   &42.6    &229.1& \textbf{46.3} \ \  &56.4 \ \  &74.0\\
\\
$^{28}$Si$K^{-}$ &$1s$&231.6   &322.4 &\textbf{90.8} \ \ &100.2\ \ &8.2    &322.2& \textbf{90.6} \ \ &99.6 \ \   &17.8    &321.8& \textbf{90.2} \ \  &98.7 \ \  &32.6\\
                 &$1p$&        &289.0 &\textbf{56.9} \ \ &68.4 \ \ &20.3    &288.2& \textbf{56.6}\ \ &67.2 \ \   &40.7    &287.0& \textbf{55.4} \ \  &65.7 \ \  & 68.6\\
\\
$^{32}$S$K^{-} $ &$1s$&261.6   &358.1 &\textbf{96.5} \ \ &107.8\ \ &14.2   &358.1& \textbf{96.5} \ \ &106.8\ \    &27.0   &357.8& \textbf{96.2} \ \  &104.1\ \  &37.0\\
                 &$1p$&        &320.1 &\textbf{58.5} \ \ &67.5 \ \ &20.2    &319.8& \textbf{58.2}\ \ &66.9 \ \    &40.6    &317.3& \textbf{55.7}\ \  &66.3 \ \  &68.8\\
\\
$^{36}$Ar$K^{-}$ &$1s$&295.2   &387.8 &\textbf{92.6} \ \ &102.4\ \ & 9.2   &387.7& \textbf{92.5} \ \ &101.3\ \   &14.4   &387.3& \textbf{92.1}  \ \  &100.0\ \  &27.8\\
                 &$1p$&        &352.9 &\textbf{57.7} \ \ &68.1 \ \ &19.2    &352.6& \textbf{57.4}\ \  &67.4\ \    &39.1   &351.6& \textbf{56.3} \ \  &66.1 \ \  &66.0\\
\\
$^{40}$Ca$K^{-}$ &$1s$&340.6   &430.4 &\textbf{89.8} \ \ &99.0 \ \ &9.3     &430.3& \textbf{89.7}\ \  &98.0\ \    &20.1    &429.7& \textbf{89.2}\ \   &96.4\ \  &36.3\\
                 &$1p$&        &400.6 &\textbf{60.0} \ \ &69.3 \ \ &18.6    &400.4& \textbf{59.8}\ \  &68.7\ \    &37.3    &399.5& \textbf{59.0}\ \  &67.6 \ \  &62.9\\
\\
$^{44}$Ti$K^{-}$ &$1s$&365.6   &455.8 &\textbf{90.3} \ \ &98.5 \ \ &9.4     &455.5& \textbf{90.0}\ \  &97.7\ \    &19.9    &454.7& \textbf{89.2}\ \  &96.3 \ \   &35.7\\
                 &$1p$&        &428.8 &\textbf{63.6} \ \ &72.3 \ \ &18.0    &428.5& \textbf{63.0}\ \  &71.7\ \    &36.1    &427.6& \textbf{62.1}\ \  &70.7 \ \  &61.0\\
\\
 \hline \hline
\end{tabular}
\end{table}
\end{center}
\end{widetext}

\subsection{Single-particle energies and widths }

The single-particle $K^-$ binding energies, $B^{s,p}_{K^-}$, the
binding energy of $K^-$, $B_{K^{-}}$, the total binding energy of
the kaonic nucleus, $B(^{A}ZK^{-})$, the total binding energy of
the ordinary nucleus, $B(^{A}Z)$ and the widths, $\Gamma$ are
listed in Tab.\ \ref{Calll}. From the Tab. we find that the
imaginary potentials (namely, the width) have a few effects on the
values of $B^{s,p}_{K^-}$, $B_{K^{-}}$ and $B(^{A}ZK^{-})$. These
values decrease with the increment of the imaginary potential
depth. The decreased values are about $1\sim 2$ MeV, if $V_0$
changes from $15$ MeV to 50 MeV. The single-particle $K^-$ binding
energies of 1s states are about $7\sim 12$ MeV larger than the
$K^-$ binding energies of 1s states, $B^{1s}_{K^{-}}$. On the
other hand, the single-particle $K^-$ binding energies of 1p
states are about $7\sim 10$ MeV larger than the $K^-$ binding
energies of 1p states, $B^{1p}_{K^{-}}$. Mare\v{s} \emph{et al.}
also predicted $B^{s,p}_{K^-}> B_{K^{-}}$ in their calculations,
they defined the difference as ``rearrangement energy", which
relates the polarization of the nuclear core by the $K^-$. The
rearrangement energies of 1s states decrease monotonically with
the nucleon number as a whole in our calculations.

The 1s state binding energies of $K^-$, $B^{1s}_{K^{-}}$, for the
listed nuclei are in the range of $B^{1s}_{K^-}= 73\sim 96$ MeV. And
the binding energies of $K^-$ for 1p states, $B^{1p}_{K^{-}}$, range
from 22 MeV to 63 MeV, increase monotonically with the nucleon
number $A$. Recently the FINUDA experiment predicted there is a
$K^-pp$ cluster with binding energy $B_{\mathrm{Kpp}}\sim 115$
MeV\cite{b4}, subtracting the binding energy between the two
protons, $B_{pp}=27.2$ MeV, the $K^-$ binding energy is $\sim 88$
MeV. Our predictions ($73\sim 96$ MeV ) are compatible with the
FINUDA experimental values  and the theoretical calculations in
\cite{a9,a10}.

The separations between the bound state 1p and 1s,
($B^{1s}_{K^{-}}-B^{1p}_{K^{-}}$), are on the order of $27\sim 62$
MeV, decrease with the increment of the nucleon number in general.
Increasing the imaginary potential depth from $V_{0}=15$ MeV to
$50$ MeV, the separations between the 1p and 1s states will
increase $\sim 1$ MeV.

On the other hand, from Tab.\ \ref{Calll}, we find that the widths
of the 1s kaonic bound states are about $12\pm 4$ MeV, $23\pm 9$
MeV and $42\pm 14$ MeV for $V_{0}=15$, $30$ and $50$ MeV,
respectively. The 1p state widths are on the order of 20, 40, for
$V_{0}=15$ and $30$ MeV, respectively. And for $V_{0}=50$ MeV, the
1p state widths are about $60\sim 80$ MeV, decrease with the
increment of the nucleon number in general. Mare\v{s} \emph{et
al.} analyzed the FINUDA experiment, considering the effects of
Fermi-motion on the decay widths the $K^-pp$ decay width $\Gamma
\sim 67$ MeV will be reduced down to 53 MeV. This result agrees
with our predictions $\Gamma^{1s}=42\pm 14$ MeV with  $V_{0}=50$
MeV.

From Tab.\ \ref{Calll}, we can also see that, if the depth
$V_{0}\leq 30$ MeV, the sum of the half widths of the 1s and 1p
states predicted by us are narrower than the separations between
$B^{1s}_{K^{-}}$ and $B^{1p}_{K^{-}}$, which implies that some
discrete states should be identified in experiment for these nuclei.
However, if $V_{0}\geq 50$ MeV the sum of the half widths of the 1s
and 1p states for all the listed nuclei, except $^{12}$C$K^-$,
predicted by us are larger than the separations of the two lowest
energy levels, thus no discrete states can be identified in
experiment.

\begin{widetext}
\begin{center}
\begin{table}[ht]
 \caption{The r.m.s. radii of neutron, proton and charge distributions, $r_n$, $r_p$
and $r_{ch}$ (in fm), respectively. $V_0$ (in MeV) is the
imaginary potential depth at normal nuclear density.} \label{Cal}
 \begin{tabular}{ccccccccccccccccccccc}  \hline \hline
\\ %\cline{3-6}
&$V_0$&$r_{p}$&$r_n$& $r_{ch}$  & & &$V_0$&$r_{p}$&$r_n$&
$r_{ch}$ & & &$V_0$&$r_{p}$&$r_n$& $r_{ch}$  \\
\hline \hline
 \\
$^{12}$C &         &\ \ 2.32 &\ \ 2.30  &\ \ 2.46               && $^{24}$Mg&         &\ \ 2.86 &\ \ 2.82 &\ \ 2.97          & &$^{36}$Ar &     &\ \ 3.26 &\ \ 3.21 &\ \ 3.36     \\
$^{12}$C$K^-$ & 15 &\ \ 2.20 &\ \ 2.18  &\ \ 2.35            && $^{24}$Mg$K^{-}$&15   &\ \ 2.80 &\ \ 2.77 &\ \ 2.92     & &$^{36}$Ar$K^{-}$ &15 &\ \ 3.22 &\ \ 3.17 &\ \ 3.32   \\
              & 30 &\ \ 2.20 &\ \ 2.18  &\ \ 2.35            &&                 &30   &\ \ 2.80 &\ \ 2.77 &\ \ 2.92                      & &&30 &\ \ 3.22 &\ \ 3.17 &\ \ 3.32 \\
              & 50 &\ \ 2.21 &\ \ 2.19  &\ \ 2.35            &&                 &50   &\ \ 2.80 &\ \ 2.77 &\ \ 2.92                      & &&50 &\ \ 3.22 &\ \ 3.18 &\ \ 3.33  \\
\\
$^{16}$O&          &2.58 &2.55 &2.70             &&$^{28}$Si&            &\ \ 2.93 &\ \  2.90  &\ \ 3.04       & &$^{40}$Ca &          &\ \ 3.36&\ \ 3.31&\ \ 3.46 \\
$^{16}$O$K^{-}$   &15&2.52 &2.49 &2.65             && $^{28}$Si$K^{-}$&15&\ \ 2.88 &\ \  2.85  &\ \ 2.99       & &$^{40}$Ca$K^{-}$&15  &\ \ 3.32&\ \ 3.27&\ \ 3.42    \\
                 &30&2.51 & 2.49 & 2.64           &&                 &30 &\ \ 2.88 &\ \  2.85  &\ \ 2.99         & &&30                &\ \ 3.32&\ \ 3.28&\ \ 3.42 \\
                 &50&2.51 & 2.48 & 2.64           &&                 &50 &\ \ 2.88 &\ \  2.85  &\ \ 2.99         & &&50                &\ \ 3.32&\ \ 3.28&\ \ 3.43  \\
\\
         $^{20}$Ne&&2.82&2.74&2.94               & & $^{32}$S &          &\ \ 3.13 &\ \ 3.09  &\ \ 3.24           & &$^{44}$Ti &         &\ \ 3.44 &\ \ 3.39 &\ \ 3.54     \\
$^{20}$Ne$K^{-}$&15&2.77&2.69&2.89               & & $^{32}$S$K^{-}$&15  &\ \ 3.08 &\ \ 3.04  &\ \ 3.19           & &$^{44}$Ti$K^{-}$&15 &\ \ 3.41 &\ \ 3.36 &\ \ 3.50      \\
              &30&2.77 & 2.68  & 2.89           & & &30                  &\ \ 3.08 &\ \ 3.04  &\ \ 3.19                         & &&30   &\ \ 3.41 &\ \ 3.36 &\ \ 3.50     \\
              &50& 2.76 & 2.68 & 2.88            & & &50                 &\ \ 3.09 &\ \ 3.04  &\ \ 3.20                      & &&50      &\ \ 3.41 &\ \ 3.36 &\ \ 3.51     \\
\\
 \hline \hline
\end{tabular}
\end{table}
\end{center}
\end{widetext}

\subsection{The static properties of the kaonic nuclei}

First, let's see Tab.\ \ref{Cal}. We find that the  r.m.s. radii of
neutron, proton and charge distributions, $r_n$, $r_p$ and $r_{ch}$
for the kaonic nuclei, $^AZK^-$, are smaller than those for the
corresponding ordinary nuclei, $^AZ$. For example, the r.m.s.\ radii
of $^{12}$C$K^-$ reduce about 0.12 fm, when a $K^-$-meson is
injected into $^{12}$C. For the other heavier kaonic nuclei, the
decreased values of the r.m.s. radii are about $0.04\sim 0.06$ fm.
Generally, the r.m.s. radii of lighter kaonic nuclei decrease more
obviously than those of heavier kaonic nuclei. The phenomenon of the
r.m.s.\ radii decreased because of a meson or hyperon being bound in
a nucleus is called ``shrinkage'', which has been found in some
lighter $\Lambda$ hypernuclei \cite{e1} in experiment. Within the
framework of RMF, we also find the shrinkage effects exist in
$\Lambda$-, $\Theta^+$- hypernuclei \cite{e2,e3}.

From the table, we also find that the r.m.s.\ radii do not change
obviously with the imaginary potential depth. If there are
differences between the r.m.s.\ radii for different imaginary
potential depth, the differences are within 0.01 fm.

To explain the ``shrinkage effect'', we should start with the $K^-$
properties in nuclear matter and the equations of motion for kaonic
nuclei in RMF theory. Firstly, the $K^-$ meson does not identify
with nucleons and can enter the nuclear cores. Secondly, the strong
$K^-N$ attraction can produce a strong $K$ field in nuclei, which
increases the strength of the scalar field $\sigma$\ (see Eq.\
(\ref{eossimp2})) and decreases the vector field $\omega$\ [see Eq.\
(\ref{eossimp3})]. This means the attraction between nucleons become
larger and the repulse between the nucleons become weaker. Thus, the
nucleons in the nucleus are bounded more tightly, which results in
the so called ``shrinkage effect".

\begin{widetext}
\begin{center}
\begin{figure}[ht]
\centering \epsfxsize=12 cm \epsfbox{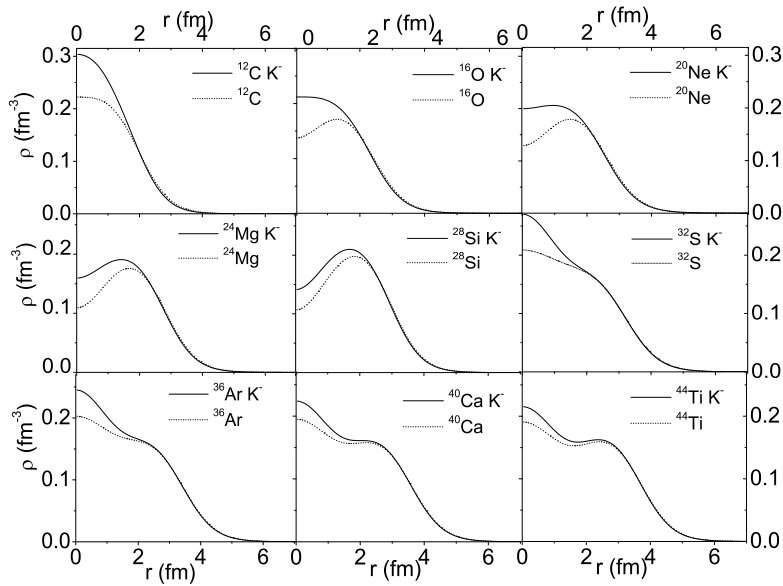}
\caption{\footnotesize Nucleon-density as a function of nucleus
radius. The solid and dotted curves are for kaonic nuclei and
corresponding ordinary nuclei, respectively.}\label{Dens}
\end{figure}
\end{center}
\end{widetext}

\subsection{nuclear density of the kaonic nuclei}

The distributions of the nuclear density for some light and moderate
kaonic nuclei in their ground states together with those of
corresponding ordinary nuclei are shown in Fig.\ \ref{Dens}. In the
calculations, we set $V_0=30$ MeV.

From the figure we can see, in the interior of the kaonic nuclei,
the nuclear densities become much denser than those of the ordinary
nuclei. The central nuclear density is enhanced by a value $0.03\sim
0.1$ fm$^{-3}$ for all the listed kaonic nuclei. In all the kaonic
nuclei,  the lowest enhanced central nuclear density is $\sim 0.03$
fm$^{-3}$ for $^{44}$Ti$K^-$, and the highest enhanced central
nuclear density is $\sim 0.1$ fm$^{-3}$ for $^{12}$C$K^-$. In all
the listed kaonic nuclei, the central nuclear density of
$^{12}$C$K^-$ is the densest, $\sim 0.31 fm^{-3}$, which is around
2.1 times the normal nuclear density $\rho_{0}$.

Our calculations give obvious enhancement of the nuclear density in
the interior of the kaonic nuclei. %The densest central nuclear
%density, 0.31 fm$^{-3}$$\sim 2.1\rho_{0}$ (for $^{12}$C$K^-$).
However, this value is much less than the prediction in  \cite{a9}.
In \cite{a9}, Akaishi \emph{et al.} gave the central density about 3
times that of the $\alpha$ particle for $^8$Be$K^-$ system, and in
\cite{a10}, the central nuclear density of kaonic nuclei gave by
Dot\'{e} \emph{et al.} even reaches to $0.71\sim1.50$ fm$^{-3}$,
which is about $4\sim10$ times the normal nuclear density with a
framework of antisymmetrized molecular dynamics. It is indeed an
amazing high central density $\rho(0)=0.81$ fm$^{-3}$$\sim5\rho_{0}$
for the many-body system $^{11}$C$K^-$ in \cite{a10} compared with
our results.

\section{Summary}

We study the kaonic nuclei within the framework of RMF theory. All
the equations are solved self-consistently. We carry out systematic
calculations for the light and moderate possible kaonic nuclei, from
$^{12}$C to $^{44}$Ti. The energy spectra of these kaonic nuclei are
studied, the ground state (1s state) binding energies of $K^-$  are
obtained, which are in the range of $73\sim 96$ MeV for all the
possible kaonic nuclei studied. The binding energies of of $K^-$ for
1p states are in the range of $22\sim 63$ MeV, and increase
monotonically with the nucleon number $A$. The separations between
the bound state 1p and 1s are in the range of $27\sim 62$ MeV, and
decrease with the increment of the nucleon number generally.

The upper limit widths of the 1s and 1p states are $42\pm 14$ MeV
and $71\pm 10$ MeV, respectively, with $V_{0}\sim 50$ MeV. And the
lower limit widths of the 1s and 1p states are $12\pm 4$ MeV and
$21\pm 3$ MeV, respectively, with $V_{0}\sim 15$ MeV.

The binding energies of $K^-$, $B_K$, will decrease $1\sim 2$ MeV,
if we increase the imaginary potential depths in their possible
range. The widths do not affect the r.m.s. radii obviously.

If the imaginary potential depth $V_{0}$ is deeper than $50$ MeV,
the sum of the half widths of the 1s and 1p states predicted by us
for $^{16}$O$K^{-}$, $^{20}$Ne$K^{-}$, $^{24}$Mg$K^{-}$,
$^{28}$Si$K^{-}$, $^{32}$S$K^{-} $, $^{36}$Ar$K^{-}$,
$^{40}$Ca$K^{-}$ and $^{44}$Ti$K^{-}$ are larger than the
separations between the 1s and 1p states for these kaonic nuclei,
which implies that no discrete states can be identified in
experiment for these nuclei. However, if $V_{0}\leq 30$ MeV the sum
of the half widths of the 1s and 1p states predicted by us are
narrower than the separations of the two lowest energy levels, thus,
the discrete $K^-$ bound states should be identified in experiment.

The shrinkage effects for kaonic nuclei are found. The interior
nucleon-densities do not increase drastically. The densest central
nuclear density predict by us is about 2.1 times the normal nuclear
density, however, which is much less than the  value predicted in
\cite{a9}.

\section*{Acknowledgements}
We would like to thank Prof. A. Gal for helpful discussions. This
work was supported in part by the Natural Science Foundation of
China (10275037, 10375074, 90203004) and China Doctoral Programme
Foundation of Institution of Higher Education (20010055012).


\begin{thebibliography}{99}
\bibitem{Kaplan}
D. B. Kaplan and A. E. Nelson, Phys. Lett. B \textbf{175}, 57
(1986).




\bibitem{brown}
G. E. Brown and C. -H. Lee \emph{et al.},
 Nucl. Phys. A\textbf{567}, 937 (1994).
%ChPT

\bibitem{c5}
G. Q. Li, C. -H. Lee, and G.E. Brown,
 Nucl. Phys. A\textbf{625}, 372 (1997).
%ChPT,RMF

\bibitem{c4}
J. Schaffner, I. N. Mishustin, and J. Bondorf,
 Nucl. Phys. A \textbf{625}, 325 (1997).
%ChPT,RMF

\bibitem{c1}
J. Schaffner, A. Gal, I.N. Mishustin, H. St\H{o}cker, and W.
Greiner,
 Phys. Lett. B \textbf{334}, 268 (1994).

\bibitem{c3}
J. Schaffner and I. N. Mishustin,
 Phys. Rev. C \textbf{53} (1996) 1416.
%RMF

\bibitem{d4}
T. Waas, N. Kaiser, and W. Weise,
 Phys. Lett. B \textbf{365}, 12 (1996);
 Phys. Lett. B \textbf{379}, 34 (1996);
N. Kaiser, P. B. Siegel, and W. Weise,
 Nucl. Phys. A\textbf{594}, 325 (1995);
W. Weise, ibid. A\textbf{610}, 35 (1996).
\bibitem{Weis} T. Waas and W. Weise, Nucl. Phys. A\textbf{625}, 287
(1997).
%Chiral

\bibitem{oset}
E. Oset, D. Cabrera, V.K. Magas, L. Roca, S.Sarkar, M.J. Vicente
Vacas and A. Ramos, PRAMANA--journal of physics, \textbf{53}, 1
(1999).
\bibitem{Hirenzaki} S. Hirenzaki, Y. Okumura, H. Toki, E. Oset, and A.
Ramos, Phys. Rev. C \textbf{61}, 055205 (2000).
\bibitem{a2}
A. Baca, C. Garc\'{i}a-Recio,
 J. Nieves, Nucl. Phys. A \textbf{673},  335 (2000).
\bibitem{a3}
A. Ciepl\'{y}, E. Friedman, A. Gal, J. Mare\v{s},
 Nucl. Phys. A \textbf{696}, 173 (2001).
%Chiral U

\bibitem{Frid}
E. Friedman, A. Gal, C.J. Batty,
 Phys. Lett. B \textbf{308}, 6 (1993);
E. Friedman, A. Gal, C.J. Batty,
 Nucl. Phys. A \textbf{579}, 518 (1994).

\bibitem{Frid1999}
E. Friedman, A. Gal, J. Mare\v{s}, A. Ciepl\'{y},
 Phys. Rev. C \textbf{60}, 024314 (1999).
%DD model
\bibitem{deepbound} E. Friedman and A. Gal, Phys. Lett. B \textbf{459}, 43 (1999).

\bibitem{a4}
Tadafumi Kishimoto,
 Phys. Rev. Lett. \textbf{83},  4701 (1999).

%kaonic nuclei

\bibitem{a6}
M. Iwasaki, K. Itahashi, A. Miyajima, H. Outa, Y. Akaishi, T. Yamazaki,
 Nucl. Instru. Meth. Phys. Res. A \textbf{473}, 286 (2001).

%\bibitem{a7}
%A. Cieply, E. Friedman, A.Gal, J. Mare\v{s},
% Nucl. Phys. A \textbf{696},173 (2001).

\bibitem{a8}
T. Yamazaki and Y. Akaishi,
 Phys. Lett. B \textbf{535}, 70 (2002).

\bibitem{a9}
Y. Akaishi and T. Yamazaki,
 Phys. Rev. C \textbf{65}, 044005 (2002).

\bibitem{a10}
A. Dot\'{e}, H. Horiuchi, Y. Akaishi, and T. Yamazaki,
 Phys. Rev. C \textbf{70}, 044313 (2004).

\bibitem{a11}
%Akinobu Dot\'{e}, Hisashi Horiuchi, Yoshinori Akaishi, Toshimitsu Yamazaki,
A. Dot\'{e}, H. Horiuchi, Y. Akaishi, T. Yamazaki,
Phys. Lett. B \textbf{590}, 51 (2004).


%Theory
\bibitem{d}
J. Mare\v{s}, E. Friedman, and A. Gal,
 Phys. Lett. B \textbf{606}, 295 (2005).
\bibitem{Galaa} J. Mare\v{s}, E. Friedman, and A. Gal, nucl-th/0601009.

\bibitem{Agnello} N. Agnello, G. Beer , L. Benussi, \emph{et al.}, Nucl. Phys. A \textbf{752}, 139c (2005).
\bibitem{Dote} A. Dote, Y. Akaishi, T. Yamazaki,  Nucl. Phys. A \textbf{754}, 391c (2005).
\bibitem{Yamagata} J. Yamagata , H. Nagahiro , Y. Okumura \emph{et
al.}, Prog. Theor. Phys. \textbf{114}, 301 (2005).
\bibitem{OsetToki} E. Oset, H. Toki, nucl-th/0509048.



\bibitem{b1}
T. Suzuki, H. Bhang \emph{et al.},
 Phys. Lett. B \textbf{597}, 263 (2004).

\bibitem{b2}
T. Suzuki, H. Bhang \emph{et al.},
 Nucl. Phys. A \textbf{754}, 375c (2005).

\bibitem{b3}
T. Kishimoto, T. Hayakawa \emph{et al.},
 Nucl. Phys. A \textbf{754}, 383c  (2005).

\bibitem{b4}
M. Agnello, G. Beer \emph{et al.} [FINUDA Collaboration],
 Phys. Rev. Lett. \textbf{94}, 212303 (2005).

\bibitem{b5} V.K. Magas, E. Oset, A. Ramos and H. Toki,
nucl-th/0601013.

\bibitem{c2}
R. Knorren, M. Prakash and P. J. Ellis,
 Phys. Rev. C \textbf{52},3470 (1995).





% \bibitem{c5}
%G. Q. Li, C.-H. Lee and G.E. Brown,
% Nucl. Phys. A \textbf{625}, 372  (1997).

%\bibitem{c6}
%E. Friedman, A. Gal, J. Mare\v{s}, and A. Ciepl\'{y},
 %Phys. Rev. C \textbf{60}, 024314 (1999).

\bibitem{c7}
X. H. Zhong, L. Li, C. H. Cai, and P.Z. Ning, Commun. Theor. Phys.
\textbf{41}, 573 (2004).



\bibitem{d0}
R. B\H{u}ttgen, K. Holinde, A. M\H{u}ller-Groeling, J. Speth, and P. Wyborny,
 Nucl. Phys. A\textbf{506}, 586 (1990);
A. M\H{u}ller-Groeling, K. Holinde, and J. Speth,
 \emph{ibid}., A\textbf{513}, 557 (1990).

\bibitem{d1}
B. D. Serot and J. D. Walecka,
 Adv. Nucl. Phys. \textbf{16}, 1 (1986).

\bibitem{d2}
P.-G. Reinhard,
 Rep. Prog. Phys. \textbf{52}, 439 (1989).

\bibitem{d3}
M. M. Sharma and M. A. Nagaragian,
 Phys. Lett. B \textbf{312}, 377 (1993).



\bibitem{d5}
%Junko Yamagata, Hideko Nagahiro, Yuko Okumura and Satoru Hirenzaki,
J. Yamagata, H. Nagahiro, Y. Okumura, and S. Hirenzaki,
nucl-th/0503039.

\bibitem{e1}
K. Tanida et al.,
 Phys. Rev. Lett. \textbf{86}, 1982 (2001).

\bibitem{e2}
Y. H. Tan, X. H. Zhong, C. H. Cai, and P. Z. Ning,
 Phys. Rev. C \textbf{70}, 054306 (2004).

\bibitem{e3}
X. H. Zhong, Y. H. Tan, G. X. Peng, L. Li, and P. Z. Ning,
Phys. Rev. C \textbf{71}, 015206 (2005).

\bibitem{Sch2006} W. Scheinast \emph{et al.}, Phys. Rev. Lett. \textbf{96}, 072301 (2006).


\bibitem{Ludh} K. Tsushima, D.H. Lu, A.W. Thomas and K. Saito, Phys.
Lett. B \textbf{443}, 26 (1998).

\bibitem{01im} A. Ramos, S. Hirenzaki, S.S. Kamalov , T.T.S. Kuo, Y. Okumura, E. Oset, A. Polls, H. Toki, L.
Tolos, Nucl. Phys. A\textbf{691}, 258 (2001).

\bibitem{chann} C. Vander Velde-Wilquet, J. Sacton, J.H. Wickens, D.N. Tovee, D.H.
Davis, Nuovo Cimento A \textbf{39}, 538 (1977).

\end{thebibliography}
\end{document}